\documentclass[aps,prl,showpacs,reprint,superscriptaddress]{revtex4-1}
\usepackage{graphicx}
\usepackage{bm}
\usepackage{amsmath}

\begin{document}

\title{Molecular quantum rotors in gyroscopic motion with a nonspreading rotational wavepacket}

\author{Sang Jae Yun}
\email[]{sangjae@kaist.ac.kr}
\affiliation{Department of Physics, Korea Advanced Institute of Science and Technology, Daejeon 305-701, Korea}
\author{Chang Hee Nam}
\email[]{chnam@gist.ac.kr}
\affiliation{Department of Physics, Korea Advanced Institute of Science and Technology, Daejeon 305-701, Korea}
\affiliation{Center for Relativistic Laser Science, Institute for Basic Science, Gwangju 500-712, Korea}
\affiliation{Department of Physics and Photon Science, Gwangju Institute of Science and Technology, Gwangju 500-712, Korea}

\date{29 March 2013}

\begin{abstract}
We provide a way of generating and observing molecular quantum gyroscopic motion that resembles gyroscopic motion of classical rotors. After producing a nonspreading rotational wavepacket called a cogwheel state, one can generate a gyroscopic precession motion by applying an external magnetic field interacting through a rotational magnetic dipole moment. The quantum rotors, realized with linear nonparamagnetic ionic molecules trapped in an ion trap, can keep their gyroscopic motion for a long time in a collectively synchronized fashion. A Coulomb-explosion technique is suggested to observe the gyroscopic motion. Despite limited molecular species, the observation of the gyroscopic motion can be adopted as a method to measure rotational g factors of molecules. 
\end{abstract}


\maketitle

\section{INTRODUCTION}
Finding a quantum state close to a motional state of a classical body has been an stimulating task to many scientists. The most famous quantum state of this kind is the coherent state \cite{Zhang1990}, which has great similarity to the classical motion of a simple harmonic oscillator. The wavepacket of the coherent state oscillates back and forth in a quadratic potential keeping the shape of the wavepacket during the oscillatory motion, so it has a nonspreading wavepacket. As another example, classical rotor motion has been pursued to find a corresponding quantum state. Though several theoretical suggestions have been made \cite{Atkins1971, Morales1999}, rotating quantum states with a nonspreading wavepacket, called cogwheel state \cite{Lapert2011}, was reported recently. In this paper, we treat quantum rotor states in classical gyroscopic motion. After preparing the cogwheel state, one can produce gyroscopic precession motion by turning on an external magnetic field which interact through the rotational magnetic moment that rotating molecules induce. We provide specific ways of generating and observing the quantum gyroscopic motion, by using linear nonparamagnetic molecular ions trapped in an ion trap. 

\begin{figure}[b]
        \centering\includegraphics[width=3.5cm]{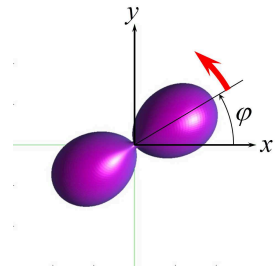}
        \caption{(Color online) An example of a quantum cogwheel state rotating in a classical sense with a nonspreading wavepacket. The figure illustrates the shape of  $\left| {\Psi _{2,\pi /6}^0} \right\rangle $ of Eq. (1). In a field-free condition, the wavepacket rotates along the z-axis with a rotating frequency of $3B$. }
\end{figure}

\section{COGWHEEL STATE}
We first survey general aspects of the cogwheel state \cite{Lapert2011} whose time evolution in a field-free condition resembles classical rotor motion. For linear molecules, the cogwheel state rotating along the z-axis is a superposition state of two rotational eigenstates $\left| {J,M} \right\rangle $, where $J$ is the angular momentum quantum number and $M$ is its z-axis projection in the laboratory frame. The cogwheel state is given by \cite{Lapert2011} 
\begin{equation}
\left| {\Psi _{n,\varphi }^J} \right\rangle  \equiv {1 \over {\sqrt 2 }}\left( {\left| {J,J} \right\rangle  + {e^{ - in\varphi }}\left| {J + n,J + n} \right\rangle } \right),
\end{equation}
where $n$ sets the number of teeth in the cogwheel-shaped rotational wavepacket and $\varphi $ is the initial angular shift of the aligned direction of the wavepacket from the x-axis. As an example, Fig. 1 shows the shape of $\left| {\Psi _{2,\pi /6}^0} \right\rangle $. In the field-free condition, the Hamiltonian is ${{\bf{H}}_0} = B{{\bf{J}}^2}$, where $B$ is the molecular rotational constant. With this free Hamiltonian, the state of Eq. (1) rotates along the positive z-axis with a rotating frequency of $B(2J + n + 1)$, and the shape of the wavepacket keeps constant in motion, meaning a nonspreading wavepacket. It is noticeable that in Eq. (1) the magnitudes of the coefficients of the two eigenstates do not need to be exactly $1/\sqrt 2 $. Regardless of the relative weights of the two eigenstates, the state still exhibits the essential features such as the nonspreading wavepacket and the rotating frequency of $B(2J + n + 1)$. Only sharpness in the shape of the `cogwheel' is degraded as the coefficients deviate from $1/\sqrt 2 $. 

In order to realize the cogwheel state of Eq. (1), molecular rotational states need to be first prepared in the state $\left| {J,J} \right\rangle $. The state preparation of $\left| {J,J} \right\rangle $ can be achieved by lowering the rotational temperature close to zero so the majority of molecules are in the rotational ground state $\left| {0,0} \right\rangle $. Rotational cooling techniques developed so far include supersonic expansion \cite{Kumarappan2006}, buffer-gas cooling \cite{Weinstein1998}, and laser cooling with trapped molecular ions in an ion trap \cite{Staanum2010, Schneider2010, Leibfried2012}. Among these techniques, we suggest only the laser cooling methods with trapped molecules to reliably generate and observe gyroscopic motion, since trapped molecules can keep their positions for a long time. Because the gyroscopic period is on the order of microseconds as explained later, it is inevitable to use trapped molecules because non-trapped diffusing molecules would collide with each other or escape the interaction region in such a time scale. Ionic molecules can be translationally cooled and trapped in an ion trap by the sympathetic cooling method with co-trapped atomic ions \cite{Ostendorf2006}. After trapping the molecules, it needs additional sophisticated techniques to cool vibrational and rotational states. For ionic polar molecules, rotational laser cooling has been demonstrated \cite{Staanum2010, Schneider2010}. In addition, for trapped ionic molecules of both polar and nonpolar types, a method which is expected to achieve the rotational ground state with a high fidelity of more than 99.9\% has been proposed \cite{Leibfried2012}. Thus, one can prepare the majority of molecules in $\left| {J,J} \right\rangle $, being particularly feasible for $J = 0$. 

\begin{figure}[tb]
        \centering\includegraphics[width=3.5cm]{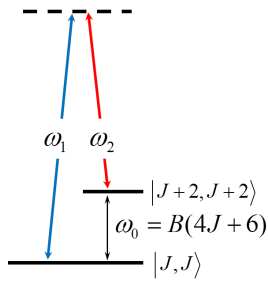}
        \caption{(Color online) Energy level diagram of two-photon rotational Raman transition between two rotational eigenstates $\left| {J,J} \right\rangle $ and $\left| {J + 2,J + 2} \right\rangle $. A pair of circularly polarized laser beams having frequencies $\omega _1^{}$ and $\omega _2^{}$ with frequency difference $\omega _0^{}$ causes the transition. }
\end{figure}
The cogwheel state of Eq. (1) with $n = 2$ can be generated by appropriate laser fields after preparing molecules in a state $\left| {J,J} \right\rangle $. The feasibleness of $n = 2$ comes from the fact that the rotational two-photon Raman transition can happen in both polar and nonpolar molecules, due to the anisotropy in their polarizability. The exclusive resonant transition between $\left| {J,J} \right\rangle $ and $\left| {J + 2,J + 2} \right\rangle $ can be performed by two circularly-polarized laser beams of which their polarization directions are counter-rotating in the x-y plane. If the frequencies of the two laser beams are in the IR range that has no resonance in vibrational or electronic transitions, the molecular rotational state involves only two-photon Raman transitions and the Hamiltonian is given by \cite{Bartels2001, Yun2012} 
\begin{equation}
{\bf{H}}(t) = B{{\bf{J}}^2} - {1 \over 2}\Delta \alpha {{\cal E}^2}(t){\cos ^2}\theta ,
\end{equation}
where $\Delta \alpha  \equiv {\alpha _\parallel } - {\alpha _ \bot }$ is the anisotropy of molecular polarizability, ${\cal E}(t)$ the laser electric field, and $\theta $ the angle between laser polarization and molecular axis. Because the energy of $\left| {J,M} \right\rangle $ is $BJ(J + 1)$, the energy gap between $\left| {J,J} \right\rangle $ and $\left| {J + 2,J + 2} \right\rangle $ is ${\omega _0} = B(4J + 6)$, as shown in Fig. 2. If the two circularly polarized beams have frequencies $\omega _1^{}$ and $\omega _2^{}$ with frequency difference $\omega _0^{}$ and equal amplitudes of ${{{{\cal E}_0}} \mathord{\left/
 {\vphantom {{{{\cal E}_0}} 2}} \right.
 \kern-\nulldelimiterspace} 2}$, the synthesized field is formed as 
\begin{gather}
\vec {\cal E}(t) = {{\cal E}_0}\cos {{\omega _1^{} + \omega _2^{}} \over 2}t \nonumber
\\
 \times \left\{ {{\bf{\hat x}}\cos \left( {{{{\omega _0}} \over 2}t + \varphi } \right) + {\bf{\hat y}}\sin \left( {{{{\omega _0}} \over 2}t + \varphi } \right)} \right\}.
\end{gather}
This is a linearly-polarized field of which polarization direction is rotating in the x-y plane with an angular frequency of ${{{\omega _0}} \mathord{\left/
 {\vphantom {{{\omega _0}} 2}} \right.
 \kern-\nulldelimiterspace} 2}$, resulting in the selection rule $\Delta J =  \pm 2$ and $\Delta M =  \pm 2$. It causes the exclusive transition between $\left| {J,J} \right\rangle $ and $\left| {J + 2,J + 2} \right\rangle $. The Raman Rabi frequency of this resonant transition is given by $\Omega  = {1 \over 4}\Delta \alpha {\cal E}_0^2\left\langle {J + 2,J} \right|{\cos ^2}\theta \left| {J,J} \right\rangle $. Thus, by properly adjusting the laser intensity and duration such that a quarter of the Rabi cycle is achieved, the superposition state of Eq. (1) with $n = 2$ is realized. For example, ${\rm{NO}}_2^ + $ molecules have $B = 12.5$ GHz and $\Delta \alpha  = 2.16 ~{{\rm{{\AA}}}^3}$ (calculated using the GAUSSIAN code \cite{Gaussian}), so the laser intensity (corresponding to ${\cal E}_0^2$) of $4.9 \times {10^6}~{\rm{W/c}}{{\rm{m}}^{\rm{2}}}$ is needed to achieve $\Omega  = 1$ MHz when $J = 0$ is chosen. 

\section{GYROSCOPIC MOTION}
With the cogwheel state produced, gyroscopic precession motion can be generated by turning on an external magnetic field. Before investigating the rotational dynamics caused by the magnetic field in detail, we first survey rotational magnetic dipole moment interacting with the magnetic field. A rotating molecule induces magnetic dipole moment because the positive ring currents from nuclei and the negative ring currents from electrons are not perfectly cancelled out. The rotational magnetic moment ${\vec \mu _r}$ is proportional to rotational angular momentum $\vec J$ with the relation ${\vec \mu _r} = {\mu _N}{g_r}\vec J$ \cite{Sutter1976}, where ${\vec \mu _r}$ is expressed in units of nuclear magneton ${\mu _N}$, and the proportionality constant ${g_r}$ is called the rotational g factor. Here we limit molecular species to nonparamagnetic molecular ions such as ${\rm{NO}}_2^ + $ or ${\rm{C}}{{\rm{H}}^{\rm{ + }}}$ whose electronic ground states are ${}^1\Sigma $ states meaning that electronic magnetic moment from both orbital and spin is zero. The requirement of nonparamagnetic molecules comes from the fact that, in the case of paramagnetic molecules, the paramagnetic moment is several thousand times bigger than the rotational magnetic moment and can cause hybridization of rotational states in a strong magnetic field forming a pendular state which results in the destruction of the cogwheel state \cite{Friedrich1992}. 

In the presence of an external magnetic field, the cogwheel state undergoes a gyroscopic precession motion while keeping the nonspreading rotational wavepacket (see \cite{Supple_Gyro} containing a movie showing the precession motion). If the magnetic field is applied along the y-axis, the Hamiltonian is given by
\begin{equation}
{\bf{H}} = B{{\bf{J}}^2} - {\mu _N}{g_r}\left| {\bf{B}} \right|{\hat J_y},
\end{equation}
where $\left| {\bf{B}} \right|$ is the magnitude of the magnetic field, and ${\hat J_y}$ is the angular momentum operator along the y-axis. Under this Hamiltonian, the precession motion arises along the y-axis and the precession frequency is given by
\begin{equation}
{\omega _p} = {\mu _N}{g_r}\left| {\bf{B}} \right|.
\end{equation}
This is because, at every time $t$ satisfying ${\mu _N}{g_r}\left| {\bf{B}} \right|t = 2\pi $, the time evolution operator ${e^{ - i{\bf{H}}t}}$ makes the quantum state the same as the previous one, except for the fast variation of the quantum phase caused by the first term of Eq. (4) which is related to the free rotation of the cogwheel state. As a specific example, ${\rm{NO}}_2^ + $ molecule has ${g_r} =  - 0.0367$ (calculated using the DALTON code \cite{Dalton2011}), so ${\omega _p}$ will be 0.28 MHz when $\left| {\bf{B}} \right| = 1$ Tesla.

\begin{figure}[tb]
        \centering\includegraphics[width=4.5cm]{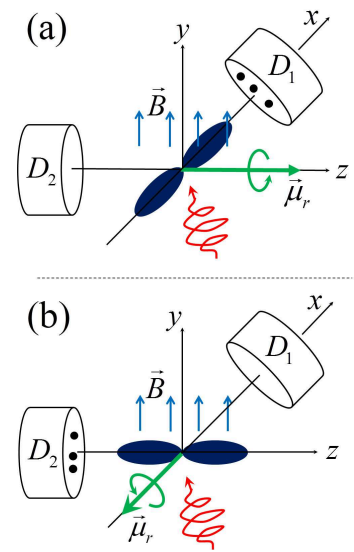}
        \caption{(Color online) Coulomb-explosion scheme for observing precession motion. After turning on the magnetic field, an intense circularly-polarized femtosecond laser pulse dissociates the molecules mainly along the aligned direction. The dissociated atomic ions are detected in the ion detectors ${D_1}$ and ${D_2}$. (a) The molecules are rotating along the z-axis meaning that the aligned direction lies in the x-y plane so that only ${D_1}$ can detect atomic ions. (b) The molecules are rotating along the x-axis, and only ${D_2}$ can detect atomic ions. }
\end{figure}
Now we provide a way of observing the molecular precession motion. Our proposed scheme exploits the fact that the cogwheel states are highly aligned to a particular direction and the aligned direction changes periodically, keeping the degree of alignment constant due to the nonspreading wavepacket. As an observation method, we suggest the Coulomb-explosion technique \cite{Dooley2003} which can directly image the rotational wavepacket. Figure 3 shows the schematic of this method. If a high-intensity circularly-polarized femtosecond laser pulse is applied to molecules, the molecules are ionized and Coulomb-exploded mainly along the aligned direction. An ion imaging detector catches dissociated atomic ions, and the aligned direction is directly obtained from the detector image. The whole sequence for observing the precession motion is as follows. After producing the cogwheel states, one turns off the laser fields and turns on the magnetic field triggering the precession motion. After a certain time delay, the ionizing femtosecond laser pulse is applied to Coulomb-explode the molecules, and the aligned direction is recorded by the ion detector. As a pump-probe scheme, the experiment is conducted repeatedly with an increased time delay. The detector image should show a periodic pattern with time delay so that the precession frequency ${\omega _p}$ can be measured. This achieves direct observation of the gyroscopic precession motion of quantum rotors. As mentioned earlier, linear nonparamagnetic ionic molecules trapped in an ion trap are used. This means that molecular ions are forming a Coulomb crystal in an ion trap. Ostendorf et al. \cite{Ostendorf2006} succeeded to trap 200 molecular ions in an ion trap using the sympathetic cooling technique. This molecular cluster in synchronized precession motion can produce a sufficient signal-to-noise ratio in an ion detector. 

Coherence time of the rotational states should be sufficient to reliably observe the gyroscopic motion which needs microsecond timescales. Schuster et al. \cite{Schuster2011} analyzed in detail the coherence time of rotational states of polar ionic molecules trapped in an ion trap. In that research, the coherence time of a single molecular ion was estimated about 0.1 s, while that of molecular cluster composed of 170000 ions was estimated about 2 ${\rm{\mu s}}$. This decreasing tendency of coherence time with increasing the number of ions is consistent with another research which reported that, when the single-qubit coherence time is $\tau $, that of N-qubit system is approximately $\tau /N$ \cite{KielpinskiThesis}. With this consideration, we can safely predict that the coherence time of rotational states of polar molecular cluster composed of 200 ions to be roughly 500 ${\rm{\mu s}}$. This is enough to observe gyroscopic motion. For nonpolar molecular ions, the situation is much better because the rotational states of nonpolar molecules are insensitive to electric field to the first order. Thus, both polar and nonpolar molecular ions trapped in an ion trap are applicable to observing their gyroscopic motion.

One potential application of the observation of the precession motion can be the measurement of rotational g factors of molecules. Provided that $\left| {\bf{B}} \right|$ is known accurately, rotational g factors ${g_r}$ is obtained from the relation ${g_r} = {\omega _p}/\left( {{\mu _N}\left| {\bf{B}} \right|} \right)$. While previous experiments to determine rotational g factors have been based on spectroscopic methods \cite{Sutter1976, Sauer1994, Amano2010JCP}, our scheme provides a new non-spectroscopic method based on the direct observation of actual precession motion for such molecular species as linear nonparamagnetic ionic molecules.

\section{CONCLUSION}
In conclusion, we have shown how one can reliably generate and observe molecular quantum gyroscopic motion, which resembles its classical counterpart, having a nonspreading wavepacket. By using linear nonparamagnetic ionic molecules sympathetically cooled and trapped in an ion trap, the molecular quantum rotors can keep their gyroscopic motion for a long time in a collectively synchronized fashion. A Coulomb-explosion technique is proposed to directly observe the gyroscopic motion. Despite limited molecular species, the observation of the gyroscopic motion in the presence of an external magnetic field will give a new non-spectroscopic method to measure rotational g factors of molecules. 

\section{acknowledgments}
We appreciate Jaewook Ahn for useful comments. This work was supported by the Ministry of Education, Science and Technology of Korea through the National Research Foundation and Institute for Basic Science.

\bibliography{Bib_total}


\end{document}